\documentstyle[psfig,12pt]{article}
\title{DECOHERENCE, EINSELECTION, AND THE EXISTENTIAL INTERPRETATION \\
(The Rough Guide)}
\author{Wojciech H. Zurek\\
Theoretical Astrophysics\\
T-6, MS B288, LANL\\
Los Alamos, New Mexico  87545\\}
\begin{document}
\maketitle

The roles of decoherence and environment-induced
superselection in the emergence of the classical from the
quantum substrate are described.  The stability of correlations
between the einselected quantum pointer states and the environment
allows them to exist almost as objectively as classical states were
once thought to exist:  There are ways of finding out what is the pointer
state of the system which utilize redundancy of their correlations with the
environment, and which leave einselected states essentially unperturbed. 
This {\it relatively objective existence\/} of certain quantum states 
facilitates operational definition of probabilities in the quantum setting.
Moreover, once there are states that `exist' and can be `found out',
a `collapse' in the traditional sense is no longer necessary --- in effect, 
it has already happened. The records of the observer will contain evidence
of an effective collapse.    The role of the preferred states in the
processing and storage of information is emphasized. 
The {\it existential interpretation} based on the relatively objective
{\it existence\/} of stable correlations between the einselected states 
of observers memory and in the outside Universe is formulated and discussed.

\section{INTRODUCTION}

The aim of the program of decoherence and einselection
(environment-induced superselection) is to describe
consequences of the ``openness'' of quantum systems to their
environments and to study emergence of the effective classicality of some 
of the quantum states and of the associated observables.  The purpose of this
paper is to assess the degree to which this program has been
successful in facilitating the interpretation of quantum theory
and to point out open issues and problems.

Much work in recent years has been devoted to the
clarification and extension of the elements of the physics of
decoherence and especially to the connections between
measurements and environment-induced
superselection.$^{1-10}$  This has included studies of emergence
of preferred states in various settings through the
implementation of predictability sieve,$^{11-14}$ refinements of
master equations and analysis of their solutions,$^{15-17}$ and study of
related ideas (such as consistent histories,$^{18-20}$
quantum trajectories, and quantum state diffusion$^{21-23}$).  A useful
counterpoint to these advances was provided by various applications, 
including quantum chaos,$^{24-26}$ einselection in the context of field
theories and in Bose-Einstein condensates,$^{27,16}$ and, 
especially, by the interplay of the original information-theoretic 
aspects$^{1,2,28}$ of the environment-induced superselection approach 
with the recent explosion of research on quantum computation$^{30-35}$ 
and related subjects$^{34-38}$. Last not least, first controlled experiment 
aimed at investigating decoherence is now in place, carried out by Brune, 
Haroche, Raimond and their collaborators$^{39}$ and additional experiments may 
soon follow as a result of theoretical$^{40-43}$ and experimental$^{44}$
developments. 

In nearly all of the  recent advances the emphasis was on specific
issues which could be addressed by detailed solutions of specific
models.  This attention to detail was necessary but may lead to impression 
that practitioners of decoherence and einselection have lost sight of their
original motivation --- the interpretation of quantum theory.  My aim
here is to sketch ``the big picture'', to relate the recent progress on
specific issues to the overall goals of the program.  I shall therefore
attempt to capture ``the whole'' (or at least large parts of it), but in
broad brush strokes.  Special attention will be paid to issues
such as the implications of decoherence for the origin of quantum
probabilities and to the role of information processing in the
emergence of `objective existence' which significantly reduces 
and perhaps even eliminates the role of the ``collapse'' of the state vector.

In short, we shall describe how decoherence converts quantum
entanglement into classical correlations and how these correlations
can be used by the observer for the purpose of prediction.  What will
matter is then encoded in the {\it relations\/} between states (such
as a state of the observer's memory and of the quantum systems). 
Stability of similar {\it co-relations\/} with the environment allows
observers to find out unknown quantum states without disturbing
them.  Recognition of this {\it relatively objective existence\/} of einselected
quantum states and investigation of the consequences of this phenomenon
are the principal goals of this paper. Relatively objective existence
allows for the {\it existential interpretation\/} of quantum theory. 
Reduction of the wavepacket as well as the ``collapse'' emerge as a consequence 
of the assumption that the effectively classical states, including the states 
of the observer's memory must exist over periods long compared to
decoherence time if they are to be useful as repositories of information.

It will be emphasized that while significant progress has been made since
the environment-induced superselection program was first formulated$^{1-4}$, 
much more remains to be done on several fronts which all have implications for 
the overarching question of interpretation. We can mention two such open issues 
right away: Both the formulation of the measurement problem and its resolution
through the appeal to decoherence require a Universe split into systems.
Yet, it is far from clear how one can define systems given an overall Hilbert
space ``of everything'' and the total Hamiltonian. Moreover, while the paramount
role of information has been recognised, I do not belive that it has been,
as yet, sufficiently thoroughly understood. Thus, while what follows is 
perhaps the most complete discussion of the interpretation implied by 
decoherence, it is still only a report of partial progress.

 \section{OVERVIEW OF THE PROBLEM}

When special relativity was discovered some twenty years before
quantum mechanics in its modern guise was formulated, it has in a
sense provided a model of what a new theory should be.  As a
replacement of Newtonian kinematics and dynamics, it was a
seamless extension.  In the limit of the infinite speed of light, $c \to
\infty$, equations and concepts of the old theory were smoothly
recovered.

When Bohr,$^{45}$ Heisenberg,$^{46}$ Born,$^{47}$ and Schr\"odinger$^{48}$
struggled to understand the implications of quantum theory,$^{49}$ one can 
sense that they had initially expected a similar seamless extension of classical
physics.  Indeed, in specific cases --- i.e., Bohr's correspondence,
Heisenberg's uncertainty, Ehrenfest's theorem --- such hopes were
fulfilled in the appropriate limits (i. e., large quantum numbers,
$\hbar \to 0$, etc.).  However, Schr\"odinger's wave packets did not
travel along classical trajectories (except in the special case of
the harmonic oscillator).  Instead, they developed into delocalized,
nonclassical superpositions.  And the tempting $\hbar \to 0$ limit did
not allow for the recovery of classical locality --- it did not even
exist, as the typical expression appearing in wavefunctions such as
$\exp (ixp/\hbar)$ is not even analytic as $\hbar \to 0$.

The culprit which made it impossible to recover classicality as a
limiting case of quantum theory was at the very foundation of
the quantum edifice:  It was the quantum principle of superposition.  It
guarantees that any superposition of states is a legal quantum state. 
This introduced a whole Hilbert space ${\cal H}$ of possibilities while only a
small fraction of states in ${\cal H}$ can be associated with the classically
allowed states, and superpositions of such  states are typically
flagrantly nonclassical.  Moreover, the number of possible
nonclassical states in the Hilbert space increases exponentially
with its dimensionality, while the number of classical states
increases only linearly.  This divergence (which is perhaps the key of the
ingredients responsible for the exponential speedup of 
quantum computations$^{30-35}$) is a measure of the
problem.  Moreover, it can be argued that it is actually exacerbated
in the $\hbar \to 0$ limit, as the dimensionality of the Hilbert space
(of say, a particle in a confined phase space) increases with
$1/\hbar$ to some power.

The first resolution (championed by Bohr$^{45}$) was to outlaw ``by fiat'' the
use of quantum theory for the objects which were classical.  This
Copenhagen Interpretation (CI) had several flaws:  It would have
forced quantum theory to depend on classical physics for its very
existence.  It would have also meant that neither quantum nor
classical theory were universal.  Moreover, the boundary between
them was never clearly delineated (and, according to Bohr, had to be
``movable'' depending on the whims of the observer). Last not least,
with the collapse looming on the quantum-classical border, there was 
little chance for a seemless extension.

By contrast, Everett's ``Many Worlds'' Interpretation$^{50}$ (MWI)
refused to draw a quantum-classical boundary.  Superposition
principle was the ``law of the land'' for the Universe as a whole. 
Branching wave functions described alternatives, all of which were
realized in the deterministic evolution of the universal state vector.

The advantage of the original Everett's vision was to reinstate
quantum theory as a key tool in search for its own interpretation. 
The disadvantages (which were realized only some years later, after
the original proposal became known more widely) include (i)~the
ambiguity of what constitutes the ``branches'' (i.e., specification
which of the states in the Hilbert spaces containing all of the conceivable
superpositions are classically ``legal'') and (ii)~re-emergence 
of the questions about the origin of probabilities (i.e., the derivation 
of the Born's formula).  Moreover, (iii)~it was never clear how to 
reconcile unique experiences of observers with the multitude of alternatives 
present in the MWI wave function.

\section{DECOHERENCE AND EINSELECTION}

Decoherence is a process of continuous measurement-like interaction between 
the system and an (external or internal) environment.  Its effect is to 
invalidate the superposition principle in the Hilbert space of an open system.  
It leads to very different stability properties for various pure states.  
Interaction with the environment destroys
the vast majority of the superpositions quickly, and --- in the case
of macroscopic objects --- almost instantaneously.  This leads of
negative selection which in effect bars most of the states and
results in singling out of a preferred stable subset of the einselected
pointer states.

Correlations are both the cause of decoherence and the criterion
used to evaluate the stability of the states.  Environment correlates
(or, rather, becomes entangled) with the observables of the system
while ``monitoring'' them.  Moreover, stability of the correlations
between the states of the system monitored by their environment
and of some other ``recording'' system (i.e., an apparatus or a
memory of an observer) is a criterion of the ``reality'' of these states. 
Hence, we shall often talk about {\it relatively objective existence\/}
of states to emphasize that they are really defined only through
their correlations with the state of the other systems, as well as to
remind the reader that these states will never be quite as ``rock
solid'' as classical states of a stone or a planet were (once) thought to be.

Transfer of a single bit of information is a single ``unit of correlation,''
whether in communication, decoherence, or in
measurement.\footnote{It is no accident that the setups
used in modern treatments of quantum communication 
channels$^{34-36}$ bear an eerie resemblance to the by now ``old
hat'' system-apparatus-environment ``trio'' used in the early
discussions of environment-induced superselection$^{1,2}$.  The apparatus
${\cal A}$ is a member of this trio which is supposed to preserve ---
in the preferred pointer basis --- the correlation with the state of the
system ${\cal S}$ with which it is initially entangled.  Hence, ${\cal A}$
is a ``communication channel.''}
It suffices to turn a unit of {\it quantum\/} correlation (i.e.,
entanglement, which can be established in course of the
(pre--)measurement -- like an interaction between two one-bit systems)
into a {\it classical\/} correlation.

This process is illustrated in Fig.~1 with a ``bit by bit''
measurement$^1$ --- a quantum controlled-not (or a {\tt c-not}).  An
identical {\tt c-not} controlled by the previously passive ``target'' bit (which
played the role of the apparatus pointer in course of the initial
correlation, Fig.~1a) and a bit ``somewhere in the environment''
represents the process of decoherence.  Now, however, the former
apparatus (target) bit plays a role of the control.  As a result, a pure
state of the system
\begin{equation}
\left|\sigma\right> = \alpha \left| 0\right> + \beta \left| 1\right>
\end{equation}
is ``communicated'' by first influencing the state of the apparatus;
\begin{equation}
\left|\Phi_{{\cal S A}} (0)\right> =
\left|\sigma\right>_{\cal S} \left|0\right>_{\cal A} \rightarrow
\alpha \left| 00\right>_{\cal SA} + \beta \left| 11\right>_{\cal SA} =
\left| \Phi_{{\cal S A}} (t_1)\right>\;,
\end{equation}
and then by spreading that influence to the environment:
\begin{equation}
\left|\Psi_{{\cal SAE}} (t_1)\right> =
\left(\alpha \left| 00\right> + \beta \left| 11\right>\right) \left|
0\right> \rightarrow
\alpha \left| 000\right> + \beta \left| 111\right> =
\left|\Psi_{{\cal S A E}} (t_2)\right>\;.
\end{equation}
Above, we have dropped the indices ${\cal SAE}$ for individual qubits (which 
would have appeared in the obvious order).

After the environment is traced out, only the correlation with
the pointer basis of the apparatus (i.e., the basis in which the
apparatus acts as a control) will survive$^{1-3,28}$:
\begin{equation}
\rho_{{\cal SA}} (t_2) = \left|\alpha\right|^2 \left|00 \right> \left< 00\right|
+ |\beta|^2 \left| 11 \right> \left< 11\right|
\end{equation}
Thus, the apparatus plays the role of the communication channel
(memory) (i)~through its ability to retain correlations with the
measured system, but also, (ii)~by ``broadcasting'' of these
correlations into the environment which is the source of decoherence
(see Fig. 1b).  Such broadcasting of quantum correlations makes them --- and 
the observables involved in broadcasting --- effectively classical.$^{29}$

The ability to retain correlations is the defining characteristic of the
preferred ``pointer'' basis of the apparatus.  In simple models of
measurement {\it cum\/} decoherence, the selection of the preferred
basis of the apparatus can be directly tied to the form of the interaction 
with the environment.  Thus, an observable $\hat O$ which commutes 
with the complete (i.e., self-, plus the interaction with the
environment) Hamiltonian of the apparatus:
\begin{equation}
\left[\hat H_{{\cal A}} + \hat H_{{\cal A E}}, \hat O\right] = 0
\end{equation}
will be the pointer observable.  This criterion can be fulfilled only in
the simplest cases:  Typically, $\left[\hat H_{{\cal A}}, \hat H_{{\cal A
E}}\right] \not= 0$, hence Eq.~(5) cannot be satisfied
exactly.

In more realistic situations one must therefore rely on more general criteria 
to which we have alluded above.  One can start by noting that the einselected
pointer basis is best at retaining correlations with the external
stable states (such as pointer states of other apparatus or record states
of the observers).  The predictability sieve$^{11-14}$ is a convenient
strategy to look for such states.  It retains pure states which
produce least entropy over a period of time long compared to the
decoherence timescale.  Such states avoid entanglement with the environment
and, thus, can preserve correlations with the similarly selected states of other
systems.  In effect, predictability sieve can be regarded as a strategy
to select stable correlations.

A defining characteristic of {\it reality} of a state is the possibility of
finding out what it is and yet leaving it unperturbed.  This criterion of
{\it objective existence} is of course satisfied in classical physics.  It
can be formulated operationally by devising a strategy which would
let an observer previously unaware of the state find out what it is
and later verifying that the state was (i)~correctly identified, and
(ii)~not perturbed.  In quantum theory, this is not possible to accomplish
with an {\it isolated\/} system.  Unless the observer knows in advance what
observables commute with the state of the system, he will in general
end up re-preparing the system through a measurement employing
``his'' observables.  This would violate condition (ii) above.  So --- it is
said --- quantum states do not {\it exist objectively}, since it is
impossible to find out what they are without, at the same time,
``remolding them'' with the questions posed by the measurement.$^{51}$

Einselection allows states of open quantum system to pass the
``existence test'' in several ways.  The observer can, for
example, measure properties of the Hamiltonian which generates
evolution of the system and of the environment.  Einselection
determines that pointer states will appear on the diagonal of the density matrix
of the system.  Hence, the observer can know beforehand what
(limited) set of observables can be measured with impunity.  He will
be able to select measurement observables that are already
monitored by the environment.  Using a set of observables 
co-diagonal in the Hilbert space of the system with the einselected
states he can then perform a nondemolition measurement to find out
what is the state without perturbing it.

A somewhat indirect strategy which also works involves monitoring
the environment and using a {\it fraction\/} of its state to infer the
state of the system.  This may not be always feasible, but this
strategy is worth noting since it is the one universally employed by
us, the real observers.  Photons are one of the most pervasive
environments.  We gather most of our information by intercepting a
small fraction of that environment.  Different observers agree about
reality based on a consensus reached in this fashion. 

That such a strategy is possible can be readily understood from the
{\tt c-not} ``caricature'' of decoherence in Fig.~1.  The einselected
control observables of the system or of the apparatus
are redundantly recorded in the environment.  One can then ``read
them off'' many times (even if each read-off may entail erasure of a
part of the information from the environment) without interacting
directly with the system of interest.

It is important to emphasize that the relatively objective existence is attained
at the price of partial ignorance.  The observer should {\it not\/}
attempt to intercept {\it all\/} of the environment state (which may be
entangled with the system and, hence, could be used to redefine its
state by sufficiently outrageous measurement$^{52}$).  Objective
existence is objective only because part of the environment has
``escaped'' with the information about the state of the system and can 
continue to serve as a ``witness'' to what has happened.
It is also important that the fraction of the environment which escapes should 
not matter, except in the two limits when the observer can intercept all of 
the relevant environment (the entanglement limit), and in the case when 
the observer simply does not intercept enough (the ignorance limit).

This robustness of the preferred (einselected) observables of the system can be 
quantified through redundancy$^{28}$, in a manner reminiscent of the recent
discussions of the error correction strategies (see, e. g., Ref. 38 and 
references therein). Consider, for example, a correlated state
\begin{equation}
\left|\Psi_{\cal SE}\right> = \left(\left|0\right>_{\cal S}
\left|000\right>_{\cal E} + \left|1\right>_{\cal S}
\left|111\right>_{\cal E}\right)/\sqrt{2}
\end{equation}
which could have arisen from a sequence of three
system-environment {\tt c-not}s.  All errors afflicting individual qubits
of the environment can be classified by associating them with Pauli
matrices acting on individual qubits of the environment.  We
can now inquire about the number of errors which would destroy the
correlation between various observables of the system and the
state of the environment.  It is quite obvious that the states
$\{\left|0\right>_{\cal S}, \left|1\right>_{\cal S}\}$ are in this sense
more robustly correlated with the environment than the states
$\{\left|+\right>_{\cal S}, \left|-\right>_{\cal S}\}$ obtained by Hadamard transform:
\begin{equation}
\left|\Psi_{\cal SE}\right>  = 
\left|+\right>_{\cal S} \left(\left|000\right>_{\cal E}+
\left|111\right>_{\cal E}\right)/\sqrt{2}\\
 \  + \   \left|-\right>_{\cal S} \left(\left|000\right>_{\cal E}-
\left|111\right>_{\cal E}\right)/\sqrt{2}
\end{equation}
For, a phase flip of any of the environment bits would 
destroy the ability of the observer to infer the state of the system
in the $\{\left|+\right>_{\cal S}, \left|-\right>_{\cal S}\}$ basis.  By
contrast, a majority vote in a $\{\left|0\right>_{\cal E},
\left|1\right>_{\cal E}\}$ basis would still yield a correct answer
concerning $\{\left|0\right>_{\cal S}, \left|1\right>_{\cal S}\}$ if {\it
any\/} single error afflicted the state of the environment. 
Moreover, when there are $N$ bits in the environment, ${N \over 2} -
1$ errors can be in principle still tolerated in the
$\{\left|0\right>_{\cal S}, \left|1\right>_{\cal S}\}$ 
basis, but in the 
$\{\left|+\right>_{\cal S}, \left|-\right>_{\cal S}\}$ 
basis a simple phase flip continues to have
disastrous consequences.

When we assume (as seems reasonable) that the probability of
errors increases with the size of the environment ($N$), so that the
``specific error rate'' (i.e., probability of an error per bit of
environment per second) is fixed, it becomes clear that the stability
of pointer states is purchased at the price of the instability of their
Hadamard-Fourier conjugates.  This stabilization of certain
observables at the expense of their conjugates may be achieved either through
the deliberate amplification or as a consequence of accidental environmental
monitoring, but in any case it leads to redundancy as it was pointed
out already some time ago$^{28}$.

This redundancy may be quantified by counting the number of ``flips''
applied to individual environment qubits which ``exchange'' the
states of the environment corresponding to the two states of the
system.  Thus, we can compute the redundancy distance $d$ between the record
states of the environment in the case corresponding to the two system states
$\phi, \psi$ given by $\{\left|0\right>_{\cal S}, \left|1\right>_{\cal S}\}$ 
in the decomposition of Eq. (6):
\begin{eqnarray*}
d (\phi,\psi) = N
\end{eqnarray*}
while in the case of the complementary observable with $\phi, \psi$ given by
$\{\left|+\right>_{\cal S}, \left|-\right>_{\cal S}\}$:
\begin{eqnarray*}
d (\phi, \psi) = 1\;.
\end{eqnarray*}
Or, in general, redundancy distance
\begin{equation}
d (\phi,\psi) = \min (n_x + n_y + n_z)
\end{equation}
is the least total number of ``flips'', where $n_x, n_y$ and $n_z$ are 
the numbers of $\hat \sigma_x$, $\hat \sigma_y$, and $\hat \sigma_z$ 
operations required to convert the state
of the environment correlated with $\left|\phi\right>$, which is given,
up to normalization constant, by:
\begin{equation}
\left|{\cal E}_\phi\right> = \left<\phi | \Psi_{\cal SE}\right>
\end{equation}
with the similarly defined $\left|{\cal E}_\psi\right>$.

Redundancy defined in this manner is indeed a measure of distance, since it is 
nonnegative:
\begin{equation}
d (\phi,\psi) \geq 0\;,
\end{equation}
symmetric:
\begin{equation}
d(\phi,\psi) = d (\psi,\phi)\;,
\end{equation}
and satisfies the triangle inequality:
\begin{equation}
d(\phi,\psi) + d(\psi,\gamma) \geq d (\phi, \gamma)
\end{equation}
as the reader should be able to establish without difficulty.

In simplest models which satisfy the commutation condition, Eq.~(5),
the most predictable set of states will consist of the eigenstates of
the pointer observable $\hat O$.  They will not evolve at all and,
hence, will be perfect memory states as well as the most (trivially)
predictable classical states.  In the more general circumstances the
states which commute with $\hat H_{\cal SE}$ at one instant will be
rotated (into their superpositions) at a later instant with the evolution
generated by the self-Hamiltonian $\hat H_{\cal S}$.  Thus, a near-zero
entropy production at one instant may be
``paid for'' by an enormous entropy production rate a while later.  An
example of this situation is afforded by a harmonic oscillator, where
the dynamical evolution periodically ``swaps'' the state vector
between its position and momentum representation, and the two representations 
are related to each other by a  Fourier transformation. 
In that case the states which are most immune to decoherence in the
long run turn out to be the fixed points of the ``map'' defined by
the Fourier transformation. Gaussians are the fixed points of the 
Fourier transformation (they remain Gaussian). Hence, coherent states 
which are unchanged by the Fourier transform are favored by decoherence.$^{11-14}$

In more general circumstances entropy production may not be
minimized by an equally simple set of states, but the lessons drawn
from the two extreme examples discussed above are nevertheless
relevant.  In particular, in the case of systems dominated by the
environmental interaction, the Hamiltonian $\hat H_{{\cal SE}}$ will
have a major say in selecting the preferred basis, while in the
underdamped case of near-reversible ``Newtonian'' limit
approximately Gaussian wave packets localized in both position and
momentum will be optimally predictable, leading to the idealization
of classical trajectories.  In either case, einselection will pinpoint the 
stable set of states in the Hilbert space.  These pointer states will be stable,
but their superpositions will deteriorate into pointer state mixtures rather 
quickly --- on the decoherence timescale --- which tends to happen very much 
faster than relaxation$^3$.

This eventual diagonality of the density matrix in the einselected basis is a 
byproduct, an important symptom, but not the essence of decoherence.  
I emphasize this because diagonality of $\rho_{\cal S}$ in some basis has been 
occasionally (mis-) interpreted as a key accomplishment of decoherence.  
This is misleading.  Any density matrix is diagonal in some basis.  
This has little bearing on the interpretation. Well-known examples of such 
accidental diagonality are the unit density matrix (which is diagonal
in every basis) and the situation where $\rho_{A\cup B} = p\rho_A +
(1 - p) \rho_B$ describes a union of two ensembles $A$ and $B$ with
density matrices $\rho_A$ and $\rho_B$ which are not co-diagonal
(i.e., $[\rho_A, \rho_B] \not= 0$).  In either of these two cases states which
are on the diagonal of $\rho_{A\cup B}$ are in effect a mathematical
accident and have nothing to do with the physical reality.

Einselection chooses preferred basis in the Hilbert space in
recognition of its predictability.  That basis will be determined by
the dynamics of the open system in the presence of environmental
monitoring.  It will often turn out that it is overcomplete.  Its
states may not be orthogonal, and, hence, they would never follow
from the diagonalization of the density matrix.  

Einselection guarantees that only those ensembles which consist of a mixture of
pointer states can truly ``exist'' in the quasi-classical sense.  That
is, individual members of such ensembles are already immune to the
measurement of pointer observables.  These remarks cannot be
made about an arbitrary basis which happens to diagonalize $\rho$ but
are absolutely essential if the quantum system is to be regarded as
effectively classical.

It is useful to contrast decoherence with the more familiar
consequence of interactions with the environment --- the noise. 
Idealized decoherence [e.g., the case of Eq.~(5)] has absolutely no
effect on the observable of interest.  It is caused by the environment
carrying out a continuous ``nondemolition measurement'' on the
pointer observable $\hat O$.  Thus, decoherence is caused by the
system observables effecting the environment and by the
associated transfer of information.  Decoherence is, in this sense, a
purely quantum phenomenon; information transfers have no effect on
classical states.

Noise, by contrast, is caused by the environment disturbing an
observable.  It is, of course, familiar in the classical context.  The
distinction between the two is illustrated in Fig.~1c, in the {\tt c-not}
language we have adopted previously.

Astute readers will point out that the distinction between noise
and decoherence is a function of the observable in terms of which
{\tt c-not} is implemented.  This is because a quantum {\tt c-not} is, in
contrast with its classical counterpart, a ``two-way street.'' When the
Hadamard transform, $\left|\pm\right> = (\left|0\right> \pm
\left|1\right>)/\sqrt{2}$, is carried out, control and target swap their
functions.  Thus, loosely speaking, as the information about the states
$\{\left|0\right>, \left|1\right>\}$ travels in one direction, the
information about the relative phase (which is encoded in their
Hadamard transforms) travels the other way.  Thus, in quantum gates
the direction of the information flow depends on the states which are
introduced at the input.

Typically, both noise and decoherence are present.  One can
reinterpret the predictability sieve$^{11}$ we have mentioned before as a
search for the set of states which maximizes the ``control'' role of the
system, while simultaneously minimizing its ``target'' role. 
Eigenstates of the pointer observable are a solution.  The phases
between them are a ``victim'' of decoherence and are rapidly erased
by the interaction with the environment.

\section{PROBABILITIES}

The classical textbook of Gnedenko$^{53}$ distinguishes three
ways of defining probability.  These are:

(i)~Definitions which introduce probability as a quantitative measure
of the {\it degree of certainty\/} of the observer;

(ii)~``Standard\footnote{``Classical'' is a more often used adjective. 
We shall replace it by ``standard'' to avoid confusion with the other
kind of classicality discussed here.} definitions'', which originate
from the more primitive concept of {\it equal likelihood\/} (and
which can be traced to the original applications of probability in
gambling);

(iii)~{\it Relative frequency\/} definitions, which attempt to reduce
probability to a frequency of occurrence of an event in a large
number of trials.

In the context of the interpretation of quantum theory, the last of
these three definitions has been invoked most often in attempts to
derive probabilities from the universal quantum dynamics.$^{54-57}$ 
The argument starts with an ensemble of identical systems (e. g., spin 
${1 \over 2}$ system) in a pure state and a definition of a relative frequency
operator for that ensemble.  The intended role of the relative
frequency operator was to act as quantum equivalent of a classical
``counter,'' but in effect it was always a meta-observable of the
whole ensemble, and, thus, it could not have been associated with the
outcomes of measurements of the individual members of the
ensemble.

A useful insight into  relative frequency observables is afforded by
the physically transparent example of Fahri {\it et~al.}$^{56}$  They
consider an ensemble of spin ${1 \over 2}$ ``magnets,'' all in an
identical state, aligned with some axis $\vec a$.  Relative frequency
observable along some other direction $\vec b$ would correspond to
a measurement of a deflection of the object with a known mass and
with a whole ensemble of spins attached to it by a (meta)
Stern-Gerlach apparatus with a field gradient parallel to
$\vec b$.  The angle of deflection would be then proportional to $\vec
a \cdot \vec b$, and ${1 + \vec a \cdot \vec b \over 2}$ would be an
eigenvalue of the frequency operator.  However, none of the spins
individually would be required to choose its ``answer.''

This approach is of interest, as it sheds light on the properties of
collective observables in quantum physics, but it does not lend
itself to the required role of supplying the probability interpretation 
in the MWI context.  A true ``frequency'' with a classical interpretation 
cannot be defined at a level which does not allow ``events'' --- quantum 
evolutions which lead to objectively existing states --- to be associated with
the individual members of that ensemble.  This criticism was made already, in somewhat 
different guise, by several authors (see Kent$^{57}$ and references therein).  
The problem is in part traceable to the fact that the relative frequency 
observables do not eliminate superpositions between the branches of 
the universal wave function and do not even define what these branches are.

Decoherence has obvious implications for the probability
interpretation. The reduced density matrix $\rho$ which emerges
following the interaction with the environment and a partial trace
will be always diagonal in the {\it same\/} basis of einselected pointer
states $\{\left|i\right>\}$.  These states help define elementary
``events.''  Probabilities of such events can be inferred from their
coefficients in $\rho$, which have the desired ``Born's rule'' form.

Reservations about this straightforward reasoning have been expressed.  
Zeh$^{58}$ has noted that interpreting coefficients of the diagonal elements 
of density matrix as probabilities may be circular.  Here we shall
therefore examine this problem more closely and prove operational
equivalence of two ensembles --- the original one {\bf o} associated
with the set of identical decohering systems, and an artificial one
{\bf a}, constructed to have the same density matrix $(\rho_{\bf o}
= \rho_{\bf a})$, but for a much more classical reason, which allows
for a straightforward interpretation in terms of relative frequencies. 
This will also shed a light on the sense in which the origin of quantum
probabilities can be associated with the ignorance of observers.

The density matrix alone does not provide a prescription for
constructing an ensemble.  This is in contrast with the classical
setting, where a probability distribution (i.e., in the phase space)
suffices.  However, a density matrix plus a guarantee that the
ensemble is a mixture of the pointer states does give such a
prescription.  Let us consider $\rho_{\bf o}$ along with the
einselected set of states $\{\left|i\right>\}$ which emerge as a result
of the interaction with the environment.  We consider an artificially
prepared ensemble {\bf a} with a density matrix $\rho_{\bf a}$ which
we make ``classical by construction.''  Ensemble {\bf a} consists of
systems identical to the one described by {\bf o}.  They are
continuously monitored by an appropriate measuring apparatus which
can interact with and keep records of each system in {\bf a}.

Let us first focus on the case of pure decoherence.  Then, in the
einselected basis:
\begin{equation}
\rho_{\bf o} (t = 0) = \sum_{i,j} \alpha_i^\ast \alpha_j \left|i \right>
\left< j\right| \rightarrow \sum_i \left|\alpha_i\right|^2 \left|i \right>
\left< i\right| = \rho_{\bf o} (t \gg t_D)\;,
\end{equation}
where $t_D$ is the decoherence timescale.  This very same evolution
shall occur for both $\rho_{\bf o}$ and $\rho_{\bf a}$.  We can
certainly arrange this by adjusting the interactions in the two cases. 
In particular, the (pointer) states $\{\left|i\right>\}$ shall remain
untouched by decoherence.

In the artificial case {\bf a}, the interpretation is inescapable.  Each
number of {\bf a} comes with a ``certificate'' of its state (which can be
found in the memory of the recording device).  Following the initial
measurement (which establishes the correlation and the first
record), the subsequent records will reveal a very boring ``history''
(i.e., $\left|i = 17\right>_{@t_1}$, $\left|i = 17\right>_{@t_2}$, \dots
$\left|i = 17 \right>_{@t_n}$, etc.).  Moreover, the observer --- any
observer --- can remeasure members of {\bf a} in the basis
$\{\left|i\right>\}$ and count the numbers of outcomes corresponding
to distinct states of each of $N$ members.  There is no ``collapse'' or
``branching'' and no need to invoke Born's rule.  All of the outcomes
are in principle pre-determined, as can be eventually verified by
comparing the record of the observer with the on-going record of
the monitoring carried out by the measuring devices permanently
attached to the system.  Individual systems in {\bf a} have ``certified'' 
states, counting is possible, and, hence, probability can be arrived at 
through the frequency interpretation for the density matrix
$\rho_{\bf a}$.  But, at the level of density matrices, $\rho_{\bf a}$ and
$\rho_{\bf o}$ are indistinguishable by construction.  Moreover, they
have the {\it same\/} pointer states, $\{\left|i\right>\}_{\bf o} =
\{\left|i\right>\}_{\bf a}$.  Since all the physically relevant elements
are identical for {\bf o} and {\bf a}, and since, in {\bf a}, the frequency
interpretation leads to the identification of the coefficients of
$\left|i \right> \left< i\right|$ with probabilities, the same must be true for
the eigenvalues of $\rho_{\bf o}$.

The ``ignorance'' interpretation of the probabilities in {\bf a} is also
obvious.  The state of each and every system is known, but until the
``certificate'' for a {\it specific\/} system is consulted, its state shall
remain unknown.  Similarly, each system in {\bf o} can be said to have
a state recorded by the environment, waiting to be discovered by
consulting the record dispersed between the environmental degrees
of freedom.  This statement should not be taken too far --- the
environment is only {\it entangled\/} with the system --- but it is
surprising how little difference there is between the statements one
can make about {\bf o} and {\bf a}.  In fact, there is surprisingly little
difference between this situation and the case where the system is
completely classical.  Consider the familiar Szilard's engine,$^{59}$
where the observer (Szilard's demon) makes a measurement of a
location of a classical particle.  The correlation between the particle
and the records of the demon can be undone (until or when demon's
record is copied).  Thus, ``collapse'' may not be as purely quantum as
it is usually supposed.  And information transfer is at the heart of the
issue in both classical and quantum contexts.  In any case, our goal
here has been a frequentist justification of probabilities.  And that
goal we have accomplished using a very different approach than these based on 
the frequency operator attempts to derive Born's formula put forward to
date.$^{54-57}$

To apply the strategy of the {\it standard definition\/} of probabilities 
in quantum physics, we must identify circumstances under which possibilities 
--- mutually exclusive ``events'' --- can be permuted without having any 
noticeable effect on their likelihoods.  We shall start with the decoherent
density matrix which has all of the diagonal coefficients equal:
\begin{equation}
\rho = N^{-1} \sum_{k=1}^N \left|k \right> \left< k\right| = {\bf 1}
\end{equation}
Exchanging any two $k$'s has obviously no effect on $\rho$ and,
therefore, on the possible measurement outcomes.  Thus, when we
assume that the total probability is normalized and equal to unity, a
probability of any individual outcome $\left|k\right>$ must be given
by:
\begin{equation}
Tr \rho \left|k \right> \left< k\right| = N^{-1}\;.
\end{equation}
It also follows that a probability of a combination of several ($n$)
such elementary events is
\begin{equation}
Tr \rho \left(\left| k_1 \right> \left< k_1\right| +
\left| k_2 \right> \left< k_2\right| + \cdots + \left| k_n \right> \left<
k_n\right|\right) = n/N\;.
\end{equation}
Moreover, when before the onset of decoherence the system was
described by the state vector
\begin{equation}
\left|\psi\right> = N^{-1/2} \sum_{k=1}^N  e^{i\phi_k}
\left|k\right>\;,
\end{equation}
the probabilities of the alternatives after decoherence in the basis
$\{\left|k\right>\}$ will be
\begin{equation}
p_{\left|k\right>} = \left|\left<k | \psi\right>\right|^2 = 
N^{-1} \ .
\end{equation}
However, in order to be able to add or to permute different
alternatives without any operational implications, we must have
assumed decoherence.  For, as long as $\left|\psi\right>$ is a
superposition, Eq.~(17), one can easily invent permutations which will
effect measurement outcomes.  Consider, for example,
\begin{equation}
\left|\psi\right> = \left(\left|1\right> + \left| 2\right> - \left|
3\right>\right)/\sqrt{3}
\end{equation}
A measurement could involve alternatives $\{\left| 1\right>, \left|
2\right> + \left| 3\right>\;,\; \left| 2\right> - \left| 3\right>\}$ and
would easily distinguish between the $\left|\psi\right>$ above and
the permuted:
\begin{equation}
\left|\psi'\right> = \left(\left| 3\right> + \left| 2\right> -
\left| 1\right>\right)/\sqrt{3}
\end{equation}
The difference between $\left| \psi\right>$ and $\left|\psi'\right>$ is
the relative phase.  Thus, decoherence and a preferred basis with
identical coefficients are {\it both\/} required to implement the
standard definition in the quantum context.

The case of unequal probabilities is dealt with by reducing it to (or at least
approximating it with) the case of equal probabilities.  Consider 
a density matrix of the system
\begin{equation}
\rho_{\cal S} = \sum_{k=1}^N p_k \left| k\right> \left< k\right|\;.
\end{equation}
We note that it can be regarded as an average of an equal probability
density matrix of a composite system consisting of ${\cal S}$ and
${\cal R}$.
\begin{equation}
\rho_{\cal SR} \cong \sum_{k=1}^N\, \sum_{j=n_k}^{n_{k+1}-1}
\left| k,j \right> \left< k,j\right|/M
\end{equation}
Here $M$ is the total number of states in ${\cal H}_{\cal SR}$, and the 
degeneracies $n_{k}$ are selected so that $p_k \simeq n_k/M$, i.e., 
\begin{equation}
n_k \cong \sum_{k' = 1}^k p_{k'} \cdot M\;,\quad
\sum_{k=1}^N n_k = M\;.
\end{equation}
For sufficiently large $M$ (typically $M \gg N$) ``coarse-grained''
$\tilde \rho_{\cal S}$ and $\rho_{\cal S}$ will become almost identical; 
\begin{equation}
\tilde \rho_{\cal S} 
=\lim_{{M \over N}\to \infty} \sum_{k=1}^N \sum_{j=n_k}^{n_{k+1}-1} \left<j\right|\rho_{\cal SR}\left|j\right>
= \lim_{{M \over N}\to \infty} Tr_{\cal R} \rho_{\cal SR}  
= \rho_{\cal S}
\end{equation}
One can now use the ``standard'' argument to obtain, first, the
probability interpretation for $\rho_{\cal SR}$ (based on the invariance of
$\rho_{\cal SR}$ under the permutations $(kj) \leftrightarrow (k'j')$), and
then use it [and Eq.~(16)] to deduce the probabilistic interpretation
of
$\rho_{\cal S}$.  Note that in the above sum, Eq.~(24), we did not
have to appeal to the actual numerical values of the eigenvalues of
$\rho_{\cal SR}$, but only to their equivalence under the permutations. 
Thus, we are not assuming a probabilistic interpretation of $\rho_{\cal SR}$
to derive it for $\rho_{\cal S}$. (We also note that the sum over auxilliary 
states above is, strictly speaking, not a conventional trace since the 
dimensions of subspaces traced out for distinct $k$'s will in general differ. 
For those concerned with such matters we point out that one can deal with 
subspaces of equal dimensionality providing that the ``dimension deficit''
is made up by auxilliary states which have zero probability.)

This completes the second approach to the quantum probabilities. 
Again, we have reduced the problem to counting.  This  time, it was a
count of equivalent alternatives (rather than of events).  In both of
these approaches decoherence played an important role.  In the standard
definition, decoherence got rid of the distinguishability of the
permuted configurations and einselection defined what they were.  
In the frequency interpretation einselection was essential --- it singled out 
states which were stable enough to be counted and verified.

Our last approach starts from a point of departure which does not rely
on counting.  Gnedenko was least sympathetic to the definitions of
probability as a measure of a ``degree of certainty,'' which he
regarded as a ``branch of psychology'' rather than a foundation of a
branch of mathematics.  We shall also find our attempt in this
direction least concrete of the three, but some of the steps are
nevertheless worth sketching.

Gnedenko's discomfort with the ``degree of certainty'' might have
been alleviated if he had been familiar with the paper of Cox$^{60}$
who has, in effect, derived basic formulae of the theory of probability
starting from such a seemingly subjective foundation by insisting that
the ``measure'' should be consistent with the laws of Boolean logic.

Intuitively, this is a very appealing demand.  Probability emerges
as an extension of the two-valued logic into a continuum of the
``degrees of certainty.''  The assumption that one should be able to
carry classical reasoning concerning ``events'' and get consistent
estimates of the conditional degree of certainty leads to algebraic
rules which must be followed by the measure of the degree of
certainty.

This implies that an information processing observer who employs classical 
logic states and classical memory states which do not interfere will 
be forced to adopt calculus of probabilities essentially identical
to what we have grown accustomed to.  In particular, likelihood of $c$
and $b$ (i.e., ``proposition $c\cdot b$'') will obey a multiplication
theorem:
\begin{equation}
\mu \left(c \cdot b | a\right) = \mu \left(c | b \cdot a\right)
\mu \left(b | a\right)\;.
\end{equation}
Above $\mu (b|a)$ designates a conditional likelihood of $b$ given
that $a$ is certain.  Moreover, $\mu$ should be normalized:
\begin{equation}
\mu \left(a|b\right) + \mu \left(\sim a | b\right) = 1\;,
\end{equation}
where $\sim a$ is the negation of the proposition $a$.  Finally,
likelihood of $c$ or $b \  (c\cup b)$ is:
\begin{equation}
\mu \left(c\cup b | a\right) = \mu (c | a) + \mu (b | a) -
\mu (c \cdot b|a)\;,
\end{equation}
which is the ordinary rule for the probability that at least one of two
events will occur.

In short, if classical Boolean logic is valid, then the ordinary
probability theory follows.  We are halfway through our argument, as
we have not yet established the connection between the $\mu$'s and the
state vectors.  But it is important to point out that the assumption of
the validity of Boolean logic in the derivation involving quantum
theory is nontrivial.  As was recognized by Birkhoff and
von~Neumann,$^{61}$ the distributive law $a \cdot (b\cup c) = (a
\cdot b)
\cup (a \cdot c)$ is {\it not\/} valid for quantum systems.  Without
this law, the rule for the likelihood of the logical sum of alternatives,
Eqs.~(26), (27) would not have held.  The physical culprit is quantum
interference, which, indeed, invalidates probability sum rules (as is
well appreciated in the examples such as the double slit
experiment).  Decoherence destroys interference between the
einselected states.  Thus, with decoherence, Boolean logic, and,
consequently, classical probability calculus with its sum rules are
recovered.

Once it is established that ``likelihood'' must be a measure (which, in
practice, means that $\mu$ is nonnegative, normalized, satisfies sum
rules, and that it depends only on the state of the system and on the
proposition) Gleason's theorem$^{62}$ implies that
\begin{equation}
\mu (a|b) = Tr \left(\left|a \right> \left< a\right| \rho_b\right)\;,
\end{equation}
where $\rho_b$ is a density matrix of the system, and $\left|a \right>
\left< a\right|$ is a projection operator corresponding to the
proposition
$a$.  Thus, starting from an assumption about the validity of {\it
classical\/} logic (i.e., absence of interference) we have arrived,
first, at the sum rule for probabilities and, subsequently, at the
Born's formula. 

Of the three approaches outlined in this section the two ``traditional''
are more direct and --- at least to this author --- more convincing. 
The last one is of interest more for its connection between logic and
probability than as a physically compelling derivation of probabilities.  
We have described it in that spirit. These sorts of logical considerations 
have played an important part in the motivation and the subsequent 
development of the ``consistent histories'' approach.$^{18-20}$

\section{RELATIVELY OBJECTIVE EXISTENCE: IN WHAT SENSE IS THE MOON THERE
WHEN NOBODY LOOKS?}

The subjective nature of quantum states is at the heart of the
interpretational dilemmas of quantum theory.  It seems difficult to
comprehend how quantum fuzziness could lead to the hard classical
reality of our everyday experience.  A state of a classical system
can be in principle measured without being perturbed by an observer
who knew nothing about it beforehand.  Hence, it is said that
classical physics allows states to exist objectively.  Operationally,
when observer A prepares a classical ensemble ${\bf a_c}$ and hides
the list ${\cal L}_A^c$ with the records of the state of each system in
${\bf a_c}$ from the observer B, it will be still possible for B to find
out the states of each member of ${\bf a_c}$ through a measurement,
with no {\it a~priori\/} knowledge.  To verify this, B could supply his
list ${\cal L}_B^c$ for inspection.  Classical physics allows ${\cal
L}_A^c$ and ${\cal L}_B^c$ to be always identical.  Moreover, both
lists will be the same as the new list ${\cal L}_A^{c'}$ with the
states of ${\bf a_c}$ remeasured by A to make sure that ${\bf a_c}$ was not
perturbed by B's measurements.  Indeed, it is impossible for A to
find out, just by monitoring the systems in ensemble ${\bf a_c}$, whether
some enterprising and curious B has discovered all that is to know about
${\bf a_c}$.  Measurements carried out on a classical ${\bf a_c}$ can be
accomplished without leaving an imprint.

This gedankenexperiment shall be the criterion for the ``objective
existence.''  When all of the relevant lists match, we shall take it as
operational evidence for the ``objective nature of measured states.'' 
In the case of a quantum ensemble ${\bf a_q}$ this experiment cannot
succeed {\it when it is carried out on a closed system}.  Observer A
can of course prepare his list ${\cal L}_A^q$ --- a list of Hilbert space
states of all the systems in ${\bf a_q}$.  B could attempt to discover what
these states are, but in the absence of any prior knowledge about the
observable selected by A in the preparation of ${\bf a_q}$ he will fail ---
he will reprepare members of ${\bf a_q}$ in the eigenstates of the
observables he has selected.  Hence, unless by sheer luck B has
elected to measure the same observables as A for each member of
${\bf a_q}$, ${\cal L}_A^q$ and ${\cal L}_B^q$ will not match.  Moreover,
when A remeasures the quantum ensemble using his ``old''
observables (in the Heisenberg picture, if necessary) following the
measurement carried out by B, he will discover that his new list
${\cal L}_A^{q'}$ and his old list ${\cal L}_A^q$ do not match either. 
This illustrates the sense in which states of quantum systems are
subjective --- they are inevitably shaped by measurements.  In a
closed quantum system it is impossible to just ``find out'' what the
state is.  Asking a question (choosing the to-be-measured
observable) guarantees that the answer (its eigenstate) will be
consistent with the question posed.$^{51}$

Before proceeding, we note that in the above discussions we have
used a ``shorthand'' to describe the course of events.  What was
really taking place should have been properly described in the
language of correlations.  Thus, especially in the quantum case, the
objectivity criterion concerned the correlation between a set of
several lists (${\cal L}_A^q$, ${\cal L}_B^q$, ${\cal L}_A^{q'}$) which
were presumably imprinted in effectively classical (i.e.,
einselected with the help of appropriate environment) sets of record
states.  The states of the systems in the ensemble ${\bf a_q}$ played a
role similar to the communication channels.  The operational
definition of objective existence of the state of the system hinges on
the ability of the state of the system to serve as a ``template,'' which
can remain unperturbed while it is being imprinted onto the records
of the initially ignorant observers.  States of isolated quantum
systems cannot serve this purpose --- they are too malleable! 
(Energy eigenstates are somewhat of an exception, but that is a
different story (Paz and Zurek, in preparation).)  We shall see below
that the einselected states of decohering quantum systems are less
fragile and can be used as a ``template''.  Again, we shall use a shorthand,
talking about states, while the real story is played out at the level of
multipartite {\it correlations}.  We assume the reader shall continue
to translate the shorthand into the ``full version.''

Consider ${\bf a_e}$, an ensemble of quantum systems subject to ideal
einselection caused by the interaction with the environment.  If the
systems are to retain their states in spite of decoherence, the
observer A has very little choice in the observables he can use for
preparation.  The menu of stable correlations between the states of
systems in ${\bf a_e}$ and his records is limited to these involving
einselected pointer states.  Only such states will be preserved by
the environment for periods of time long enough to contemplate the
gedankenexperiment described above.

The task of the observer B (who is trying to become correlated with
the stable states of ${\bf a_e}$ without destroying pre-existing stable
correlations established by the observer A) is simplified.  As soon as
he finds out what are the pointer observables, he can measure at
will.  He can be sure that -- in order to get sets of records with 
predictive power -- A must have selected the same pointer observables
to prepare ${\bf a_e}$.  And as soon as the pointer observables are known,
their eigenstates can be found without being perturbed.  Moreover, B
will be measuring observables which are already being monitored by
the environment, so his measurements will have no discernible effect on the
states of the systems in ${\bf a_e}$.

Hence, either A was smart enough to choose pointer states (in which
case his lists ${\cal L}_A^e$, ${\cal L}_A^{e'}$, \dots will all be 
identical) and B's spying will not be detected, {\it or\/} A chooses
to measure and prepare arbitrary states in the Hilbert space, guaranteeing their
deterioration into mixtures of pointer states at a rapid decoherence
rate.  In this second case A's lists will reflect a steady increase of
entropy caused by the mismatch between the observable he has
elected to measure and the pointer observables he should have
measured.  And B's spying will most likely still be undetected
(especially if he is smart enough to focus on the pointer
observables).

Let us now compare the three variants of the gedankenexperiment
above.  In the classical case, anyone can find out states of the
systems in ${\bf a_c}$ without perturbing them.  Prior information is
unnecessary, but only classical (i.e., localized, etc.) states can be
used.  In the case of a quantum isolated system an enormous variety
of quantum states --- including all conceivable superpositions of the
classical states --- can be prepared by A in ${\bf a_q}$.  Now B's
measurement will almost inevitably reprepare these states, unless
somehow B knows beforehand what to measure (i.e., what
observables has A measured).  In the third case --- quantum, but with
decoherence and einselection --- the choices of A are limited to the
preferred pointer states.  Only a very small subset of all the
superpositions in the Hilbert space ${\cal H}$ is available.  Moreover,
the environment is already carrying out continuous monitoring of the
observables it has elected to ``measure.''  B can use the correlations
established between the system and the state of the environment to
find out what are the preferred observables and to measure them. 
He will of course discover that his list matches A's lists and that A
could not have detected B's ``spying.''

When the states can be ``revealed'' without being reprepared in the
process, they can be thought to exist objectively.  Both the classical
case and the quantum plus einselection case share this feature.  The
environment-induced superselection simultaneously decreases
the number of states in ${\cal H}$ while allowing the einselected
states to ``exist objectively'' --- to be found out without the
necessity of repreparing them in the process. 

In fact, the measurements we carry out as observers are taking an
even more immediate advantage of the monitoring carried out by the
environment.  Our measurements are almost never direct --- nearly
without exception they rely on interception of the information
already present in the environment.  For instance, all of the visual
information we acquire comes from a tiny fraction of the photon
environment intercepted by the rod and cone cells in our eyes. 
Indeed, this is perhaps the best strategy observer B could have used
in the third version of the gedankenexperiment above.  Rather than
directly interacting with the system in ${\bf a_e}$, he could have monitored the
environment.  An imprint left in a small fraction of its state is usually
enough to determine the state of the system --- the environment
contains a vastly redundant record of the pointer observables.$^{28}$ 
Thus, perception of classical reality seems inevitable for the
observers who --- like us --- rely on the second-hand information, on
the correlations acquired indirectly, from the environment.

In a sense the environment plays the role of a commonly accessible 
``internet--like'' data base which allowes one to make copies of 
the records concerning the state of the system. There is no need to measure 
the system directly --- it suffices to consult the relevant ``web page''. 
And there is no danger of altering the state of the system:
Nonseparability and other such manifestations of quantum theory could reappear
only if, somehow, all of the widely disseminated copies of the information were
captured and processed in the appropriate (quantum) manner. The difficulty 
of such an enterprise in the macroscopic domain (which we have quantified before
by the redundancy distance, Eqs. (6)-(12)) is a measure of 
irreversibility of the decoherence--induced ``reduction of the wavepacket''.

We have just established that states of quantum systems
interacting with their environments exist much like the classical
states were presumed to exist.  They can be {\it revealed\/} by
measurements of the pointer observables which can be ascertained
without prior knowledge.  In particular, indirect
measurements --- observes monitoring the environment in search
of the imprints of the state of the system --- seem to be the
safest and at the same time most realistic way to reveal that state. 
Moreover, there are many fewer possible einselected states than
there are states in the Hilbert space.  Thus, the relative objectivity
based on the system-environment correlations and, hence, on
decoherence and einselection, comes at a price:  The choice is
severely limited.\footnote{How limited?  There are of course
infinitely many superpositions in ${\cal H}$ of finite dimensionality,
but that is already true of a spin ${1 \over 2}$ Hilbert space, and it does not
capture the reason for the incredible proliferation of
superpositions.  In the Hilbert space of a decohering system, there will
be $N \sim {\rm Dim} ({\cal H})$ pointer states $\{\left|k\right>\}$.  For
a typical superposition state $\left|\psi\right>$ composed of all $N$
states, with the possibilities ``coarse grained'' by assuming that all
$\left|\psi\right>$ have a form:
\begin{displaymath}
\left|\psi\right> = {1 \over \sqrt{N}} \sum_k (-)^{i_k}
\left|k\right>
\end{displaymath}
where $i_k$ is  0 or 1, there will be
\begin{displaymath}
{\cal W} \sim 2^N 
\end{displaymath}
such superpositions.  That is, even when we set all the coefficients to
have the same absolute value, and coarse-grain phases to the least
significant (binary) digit, we will have exponentially many
possibilities.}

In the above operational approach to the definition of existence, we
have made several simplifying assumptions.  We have (i)~neglected
the evolution; (ii)~assumed perfect decoherence, and; (iii)~focused
on observers with ``perfect knowledge,'' i.e., used pure states rather
than mixtures as initial conditions.  All of these assumptions can be
relaxed with relatively little pain.  Hamiltonian evolution {\it
alone\/} would not be a problem --- the system could be described in
Heisenberg's picture.  But the combination of evolution and
decoherence will lead to complications, resulting in a preferred
basis which is imperfect$^{12-14,17}$ --- even the optimal pointer states would
eventually deteriorate into mixtures, albeit on a timescale long
compared to the decoherence timescale for random superposition in ${\cal H}$.

This difference between the einselected states and arbitrary
superpositions could be quantified by defining the predictability
horizon:
\begin{equation}
t_p = \int\limits_0^\infty \left(H_{EQ} - H (t)\right)\,dt/
\left(H_{EQ} - H (0)\right)
\end{equation}
This characterizes the timescale over which the initial information
$H_{EQ} - H(0)$ is lost as the von~Neumann entropy, $H(\rho)= -Tr \rho \log \rho$,
approaches the long-term (equilibrium) value $H_{EQ}$.  Easier to
compute (and similarly motivated)
\begin{equation}
t'_p = \int\limits_0^\infty Tr \left(\rho_t^2 - \rho^2 (\infty)\right)\,
dt
\end{equation}
should supply equivalent information.  Thus $t'_p$ would be short (and of the 
order of the decoherence time) for a typical initial state in the Hilbert space.
By contrast, the predictability horizon may be long (and, perhaps, even
infinite) for pointer states of integrable systems, while for chaotic
systems one would anticipate predictability timescales determined
by the Lyapunov exponents when decoherence dominates.$^{24}$

The gedankenexperiment at the foundation of our ``operational
definition of existence'' is based on the comparison of records of two
observers A and B\null.  It could now be repeated, provided that the
duration of the experiment is brief compared to the predictability
timescale, or that the natural rate of information loss is accounted
for when evaluating the final results.

In fact, the predictability sieve could be implemented using
this strategy.  Einselected pure states will maximize $t_p$. 
Moreover, such procedure based on the predictability timescale can
be easily applied to compare pure and mixed states.  That is, one can
find out how much more durable are various correlations between the
observer's records of the coarse-grained measurements.  The key
difference from the original predictability sieve,$^{11}$ which has
been successfully used to demonstrate the special role of
Gaussians,$^{11-14}$ is a somewhat different sieve criterion, which
may even have certain advantages.

All these caveats and technicalities should not obscure the central
point of our discussion.  Environment-induced superselection allows
observers to anticipate what states in the Hilbert space have a
``relatively objective existence'' and can be revealed by
measurements without being simultaneously reprepared.  {\it Relatively
objective existence\/} is a deliberate double entendre, trying to
point out both the relative manner in which existence is defined (i.e., through
correlations, similar in spirit to the relative states of
Everett$^{50}$) and a reminder that the existence is not absolutely
stable but, rather, that it is purchased at the price of decoherence
and based on the monitoring by the environment.

Concerns about the predictability timescale do not imply that, on
timescales long compared to $t_p$, the states of the systems in
question do not ``exist.''  Rather, $t_p$ indicates the predictability
horizon on which evolution and decoherence destroy the relevance of
the ``old'' data (the record-state correlation).  But even then the
essence of our definition of reality --- the ability of the observer to
``reveal'' the state --- captures the essence of ``existence.''

\section{THE EXISTENTIAL INTERPRETATION}

The interpretation based on the ideas of decoherence and
einselection has not really been spelled out to date in any detail.  I
have made a few half-hearted attempts in this direction,$^{11,63}$
but, frankly, I was hoping to postpone this task,
since the ultimate questions tend to involve such ``anthropic''
attributes of the ``observership'' as ``perception,'' ``awareness,'' 
or ``consciousness,'' which, at present, cannot be modelled 
with a desirable degree of rigor.

It was my hope that one would be able to point to the fact that
decoherence and einselection allow for existence (as 
defined operationally through relative states and correlations in the
preceding section) and let those with more courage than I worry
about more esoteric matters.  I have been gradually changing my
mind as a result of penetrating
questions of my colleagues and the extrapolations (attempted by
the others) of the consequences of decoherence and einselection
which veered in the directions quite different from the one I 
have anticipated.  (See references 64--67 for a sample of questions,
criticisms, and attempts at an interpretation.)  Moreover, while there are 
``deep'' questions which may be too ambiguous to attack with the tools used by
physicists, there are aspects of information processing which bear direct
relevance for these deeper issues, and which can be profitably analysed in a
reasonably concrete setting. Here I intend to
amplify some of the points I have made before and to provide the
``next iteration'' by investigating consequences of environmental
monitoring a step or two beyond the operational definition of
``existence.'' I shall proceed in the general direction indicated
earlier,$^{11, 63}$ and focus on the stability of the einselected
correlations.

We start by noting that the relatively objective existence of certain states 
``negotiated'' with the environment has significant consequences for the 
observers and their information-processing abilities. In the gedankenexperiments
involving observers~A and~B in the preceding section
we could have equally well argued for the objective existence of the
states of their memory cells.  Again, superpositions of all the
possibilities are ruled out by einselection, and the brain of an
observer can preserve, for long periods of time, only the pointer states of 
its neurons.  These states exist in the same relatively objective sense
we have defined before --- they can be revealed (correlated with)
the states of other neurons without having to be simultaneously
re-prepared.  Indeed, real neurons are coupled very strongly to their
environments and certainly cannot exist in superpositions.
Their two stables states are characterized by different rates of
firing, each a consequence of a nonequilibrium
dissipation-dominated phenomena, which are bound to leave a very
strong imprint on the environmental degrees of freedom not
directly involved in the information processing.  In an obviously
overdamped system operating at a relatively high temperature,
the inability to maintain superpositions is not a
surprise.\footnote{Neurons work more like a diode or a transistor,
relying on two stable steady states (characterized by different firing
rates) for stability of the logical ``0'' and ``1,'' rather than the
two-state spin ${1 \over 2}$-like systems, which often end up playing
the neuron's roles in models of neuron networks invented by theoretical
physicists.  I believe, however, that for the purpose of the ensuing
discussion this distinction is not essential and I will continue to
invoke {\it licentia physica theoretica\/} and consider spin ${1 \over
2}$-like neurons for the remainder of this paper.}

When we assume, as seems reasonable, that the states of neurons are
the ``seat'' of memory and that their interactions lead to information
processing (which eventually results in ``awareness,'' and other such
``higher functions''), we have tangible hardware issues to analyze. 
Indeed, at this level of discussion there is little fundamental
difference between a brain and a massively parallel, neural
network--like, {\it effectively classical\/} computer.

The ability to process information concerning states of objects
external to memory (for, say, the purpose of prediction) is then
based on the stable existence of correlations between the record
bits of memory and the state of the object.  It is fairly easy to
imagine how such a correlation can be initially established either by a
direct measurement or, as we have noted previously, through an
indirect process involving environment.  For the reliability of
memories, it is absolutely crucial that this correlation be
immune to further correlations with the environment, i.e., to
decoherence.  Following a measurement (and the initial bout of
decoherence), the reduced joint density matrix of the system and the 
relevant part of memory and the environment will have the form:
\begin{equation}
\rho_{\cal SM} = \sum_i p_i \left|s_i \right> \left< s_i\right|
\left|\mu_i\right> \left< \mu_i\right| 
\end{equation}
The predictability horizon can be defined as before through
\begin{equation}
t_p^{(i)} = {\int\limits_p^\infty \left(H (s_i,\mu_i;t) -
H (s_i, \mu_i;\infty)\right)\, dt \over
H(s_i, \mu_i;0) - H(s_i,\mu_i; \infty)}
\end{equation}
for individual outcomes.  Here $H$ can stand for either Shannon-von~Neumann, 
or linear (or still other) measure of information content of the conditional
(re-) normalized $\left<\mu_i \left|\rho_{\cal SM}\right| \mu_i\right>$.  
After a perfect measurement there is a one-to-one correlation between the 
outcome and the record (Eq. (31)). It will, however, deteriorate with time as 
a result of dynamics and the openness of the system, even if the 
record--keeping memory states are perfectly reliable.$^{68}$ The predictability 
timescale for memory-system joint density matrices has a more 
specific interpretation than the one defined by Eqs. (29) and (30).  It is 
also safe to assume that memories use stable states to preserve records. 
In this ``no amnesia'' case, $t_p^{(i)}$ will measure the timescale on which
the acquired information is becoming a useless ``old news'' because
of the unpredictable evolution of the open system. Predictability horizon
can (and typically will) depend on the outcome.

We note that more often than not, both the states of memory and the
states of the measured systems will be mixed, coarse-grained states
inhabiting parts of large Hilbert spaces, rather than pure states.  Thus the
record $\mu_i$ will correspond to $\rho_{\mu_i}$ and $Tr
(\rho_{\mu_i}
\rho_{\mu_j}) \cong \delta_{ij}$.  It is straightforward to generalize
Eqs.~(31) and~(32) to cover this more realistic case.  Indeed, it is
likely that the memory states will be redundant, so that the likely
perturbations to the ``legal'' memory states shall retain
orthogonality.  This would allow for classical error correction, such as is
known to be implemented, for example, in neural circuits responsible
for photodetection in mammalian eyes, where several
(approximately seven) rods must fire more or less simultaneously
to result in a record of detection of a light source.

Observers will be able to make accurate predictions for as long as a
probabilistic equivalent of logical implication shall be valid, that is, as
long as the conditional probability $g(t)$ defined by:
\begin{equation}
p \left(\sigma_i (t)|\mu_i\right) =
p \left(\sigma_i (t), \mu_i\right)/p (\mu_i) = g(t)
\end{equation}
is close to unity.  Here $\sigma_i (t)$ is a ``proposition'' about the
state of the system at a time $t$.  One example of such a formal
equivalent of a proposition would be a projection operator onto a
subspace of a Hilbert space.  When $\sigma_i (t)$ is taken to be a pure 
state $\left|s_i (t)\right>$, and $t \ll t_p^{(i)}$ so that $g(t) \cong 1$,
Eq.~(33) becomes in effect a formal statement of the ``collapse''
axiom of quantum measurement theory. For, memory will continue to 
rediscover the system in the same state upon repeated re-measurements.

Again --- as in the preceding section --- relatively objective existence
of {\it correlations\/} (established in contact with the environment,
with their stability purchased at the expense of the limitation of
the possible states of memory and the measured system) is decisive
for the predictive utility of the records.  These records must be
repeatedly and reliably accessible for the other parts of memory to
allow for information processing.  This is why the record states which
{\it exist\/} (at least in the relatively objective operational sense
employed before) are essential for the reliability of memories
inscribed in open systems.  They can be re-measured by the other
memory cells spreading the useful correlation throughout the
information processing network of logical gates but suffer no ill
effects in the process. 

The record state $\left| \mu_i \right> $ must then obviously be decoherence
resistant, but the same should be true for the measured states
$\left|s_i(0)\right>$ and (hopefully) for their near-future
descendants $\left|s_i(t)\right>$.  Only then will the correlation between
memory and the state of the system be useful for the purpose of
predictions.  One can analyze this persistence of quasi-classical
correlations from various points of view, including the algorithmic
information content one.  We shall mention this idea here only briefly
as a more complete account is already available.$^{68}$  In essence,
when $\left| s_i (t)\right>$ evolves predictably, a sequence of
repeated measurements of the appropriate observables yields a
composite record $R = \{\mu_{@t_1}^{(1)}, \mu_{@t_2}^{(2)},...,
\mu_{@t_n}^{(n)}\}$, which will be all derivable from the 
initial $\mu_{@t_0}^{(0)}$ and from
the algorithm for the evolution of the monitored system.  This
predictability could be expressed from the viewpoint of the
observer by comparing the algorithmic information content of $R$
with its size in bits.  When the whole $R$ can be computed from the
initial condition, the algorithmic information content $K(R)$
(defined as the size of the shortest program for a universal
computer with the output $R$$^{69}$) is much less than the size of the
``uncompressed'' $R$ in bits.  An illustrative (if boring) example of
this would be a sequence of records of a fixed state of an object such
as a stone.  Then $R$ would simply be the same record, repeated over
and over. This is of course trivially algorithmically compressible.  

This immobility of objects such as stones is the basic
property which, when reflected in the perfectly predictable
sequence of records, provides a defining example of (the most basic
type of) perception of existence, of permanence which defines 
``classical reality''.  In this case, the same set of observables
``watched'' by the environment is also being watched by the observer.

In general, the state of a system evolving in contact with the
environment will become less and less closely correlated with its initial
state.  Therefore, entropy will increase, and the reliability of the
implication measured by the conditional probability $p(\sigma_i
(t)|\mu_i)$ will decrease.  Alternatively, one may want to retain the
safety of predictions (i.e., have $p(\sigma_i(t)|\mu_i)$ close to unity
at the price of decreased accuracy).  This could be accomplished by
choosing a safer (but less precise) prediction $\tilde \sigma_i(t)$
which includes $\sigma_i (t)$ with some ``error margin'' and thus supplies 
the requisite redundancy.

For a judiciously selected initial measurement the conditional probability
$p (\sigma_i (t)|\mu_i)$ will decrease relatively slowly (for
example, on a dynamically determined Lyapunov timescale in the
case of chaotic systems$^{68}$), while for measurements which result
in the preparation of a ``wrong'' initial condition --- a state at odds
with einselection --- the conditional probability would diminish suddenly,
on a near-instantaneous decoherence timescale.

Typically, the prediction $\sigma_i$ will not be a pure state but a suitably 
macroscopic patch in the phase space (and a corresponding ``chunk'' of 
the Hilbert space).  Increase in the size of the patch
will help extend the relative longevity of the predictive power of the
records at the expense of the accuracy.  Nevertheless, even in this
case predictive power of the old records shall eventually be lost
with time.

The memory must be stored in robust record states, which will
persist (ideally forever, or, at least, until they are deliberately
erased).  Deliberate erasure is an obvious strategy when the records
outlive their usefulness.

In this picture of a network of effectively classical memory elements 
interconnected with logical gates, stability of the records is essential.  
It is purchased at the price of ``censorship'' of the vast majority of all of 
the superpositions which are in principle available in the Hilbert space.  It
is in an obvious contrast with quantum computers,$^{30-35}$
where all of the superpositions are available for information processing.
and where the states of memory are unstable and prone to the environment-induced
decoherence and other errors.

Let us now consider how such a quasi-classical, environmentally
stable memory ``perceives'' the Universe.  To avoid generalities, we
consider --- from the point of view of this memory --- the task of
determining what are the classical branches of the universal state
vector. That is, we shall ask the observer to find out what are the 
pointer states in his branch of the Universe.

In effect, we have already presented all of the elements necessary
for the definition of the branches in the course of the discussion of
existence in the preceding section.  A branch is defined by its
predictability --- by the fact that the correlations
between its state and the records of the observer are
stable.  In other words, a branch is defined by the fact that it does
not split into branches as rapidly as a randomly selected state.

The observer is ``attached'' to the branch by the correlations between its
state and the einselected states which characterize the branch. 
Indeed, the observer is a part of his branch.  In the case of perfect
predictability (no entropy production; initial records of the observer
completely determine the future development of the branch), there
would be no further branching.  An observer will then be able --- on
the basis of his perfect predictability --- to develop a view that
the evolution of the Universe is completely deterministic and that
his measurements either confirm perfect predictability (when
carried out on the system he has already measured in the past) or
that they help reveal the pre-existing state of affairs within ``his''
branch.

This classically motivated and based on Newtonian intuitions ``single branch'' 
(single trajectory) limit of quantum physics is responsible for the
illusion that we live in a completely classical Universe.  It is an
excellent approximation in the realm of the macroscopic, but it
begins to fail as the resolution of the measurements increases.  One
example of failure is supplied by quantum measurements and, more generally,
by the situations where the state of the macroscopic object is
influenced by the state of a quantum object.  Then the past initial
condition is exhaustive --- the observer knows a pure state, i.e.,
all that is possible to know --- and yet this is demonstrably
insufficient to let him determine the future development of his
branch.  For, when the past correlation is established in a basis
incompatible either with the monitoring carried out by the
environment, or when it prepares a state which is not an eigenstate of
the new measurement, the new outcome cannot be inferred from the old
initial condition.

Relatively objective existence is at the core of the definition of
branches.  Stability of the correlations between the state of the
observer and the branch is responsible for the perception of
classicality.  Stability of the record states of the observer is an
obvious precondition.  The observer with a given set of records is
firmly attached to the branch which has induced these records --- he
``hangs on'' to the branch by the correlations.  He may even be
regarded as a part of that branch, since mutual correlations between
systems define branches.

Observers described here are quite different from the aloof
observers of classical physics, which simply ``note'' outcomes of
their measurements by adding to their abstract and immaterial
repository of information.  In a quantum universe {\it information is
physical}$^{70}$ --- there is simply {\it no information without
representation}.$^{63}$  In our context this implies that an observer
who has recorded one of the potential outcomes is {\it physically
distinct\/} from an observer who has detected an alternative
outcome.  Moreover, these two states of the observer are
objectively different --- they can be ``revealed'' from the outside
(i.e., by monitoring the environment in which record states are
immersed) without disturbing the observer's records.

The question ``why don't we perceive superpositions'' (which has been
repeated also by some of those who investigate and support ideas of
decoherence$^{71}$) has a straightforward answer.  The very
physical state of the observer and, thus, his {\it identity\/} is a
reflection of the information he has acquired.  Hence, the acquisition
of information is not some abstract, physically insignificant act, but
a cause of reshaping of the state of the observer.  An exaggerated
example is afforded by the famous case of Schr\"odinger's cat.$^{48}$ 
A cat that dies as the result of an amplified quantum event will 
certainly be easily distinguishable from the cat that lives on (and can
continue observations).  Similarly, an effectively classical
computer playing the role of an observer will be measurably distinct
depending on what outcome was recorded in its data bank.

Coherent superpositions of two memory states will disappear on the
decoherence timescale in the presence of the environment.  Hence, a
coherent superposition of two distinct identities of an observer
does not exist in the relatively objective operational sense
introduced previously.  Even in the rare cases when a memory bit of
an observer enters into a bona fide entanglement with an isolated
quantum system, decoherence will intervene and turn that
entanglement into an ordinary classical correlation in the basis
defined by the einselected record states.

The interpretation which recognizes that decoherence and 
environment-induced superselection allow for the {\it
existence\/} of states at the expense of the superposition principle
is known as the {\it existential interpretation}.$^{11,63}$  It accounts
for the inability of the observers to ``perceive'' arbitrary
superpositions.  The (relatively) objective existence of the records is a
precondition for their classical processing and, therefore, for
perception.\footnote{It is amusing to speculate that a truly quantum
observer (i.e., an observer processing quantum information in a
quantum computer-like fashion) might be able to perceive
superpositions of branches which are inaccessible to us, beings
limited in our information processing strategies to the record
states ``censored'' by einselection.}

It is easy to design logical states which would distinguish
between objectively existing records and accidental states. 
Redundancy is the key.  Thus, when an entanglement between a
(two-state) system and a memory cell develops,
\begin{equation}
\left|\phi_{{\cal S}{\mu}}\right> = \alpha \left|\uparrow\right> \left|
1\right> +
\beta \left|\downarrow\right> \left| 0\right>\;,
\end{equation}
and, under the influence of the environmental decoherence, rapidly
deteriorates to a classical correlation
\begin{equation}
\rho_{{\cal S}{\mu}} = \left|\alpha\right|^2 \left|\uparrow\right> 
\left< \uparrow\right| \left| \uparrow\right> \left< 1\right| +
\left|\beta\right|^2 \left|\downarrow\right> \left<\downarrow\right|
\left| 0 \right> \left< 0\right|\;,
\end{equation}
the reliability of the record state in an arbitrary basis can be in
principle tested by the other parts of the memory.  Repeated
measurements of the same memory cell in different bases and
comparing longevity of the state in the $\{\left|0\right>, \left|
1\right>\}$ basis with the (lack of) longevity of the state in the
$\{\left|+\right>, \left|-\right>\}$ basis would do the trick.  Thus, as a
consequence of decoherence and einselection,
\begin{eqnarray}
\rho_{\cal S M} & = &
\left|\alpha\right|^2 \left|\uparrow\right> \left<\uparrow\right|
\left| 1\right> \left< 1\right| \left|1'\right> \left< 1'\right|
\left| 1''\right> \left< 1''\right|\ldots \nonumber \\
&+ & \left|\beta\right|^2 \left|\downarrow\right>
\left<\downarrow\right|
\left| 0\right> \left< 0\right| \left|0'\right> \left< 0'\right|
\left| 0''\right> \left< 0''\right|\ldots\;,
\end{eqnarray}
in the transparent notation, while in the case of measurements of
$\{\left|+\right>, \left|-\right>\}$ record states there would be no
correlation between the consecutive measurements carried out at
intervals exceeding decoherence timescale.  Instead of two branches
of Eq.~(36), there would be $2^N$ branches, where $N$ is the number
of two-state memory cells implicated.  And a typical branch would be
algorithmically random, easily betraying unreliability of the
$\{\left|+\right>, \left|-\right>\}$ record states.

This simplistic model has no pretense to realism.  Rather, its aim is to
demonstrate a strategy for testing what is reliably known by the
observer.  A slightly more realistic model would entail redundant
records we have mentioned already, Eqs. (6)-(12).  Thus, the initial correlations
would involve several $(n)$ memory cells:
\begin{eqnarray}
\rho_{{\cal S}{\mu^n}} & = &
\left|\alpha\right|^2 \left|\uparrow\right> \left<\uparrow\right|
\left| 1\right> \left< 1\right|_1 \left|1\right> \left< 1\right|_2
\ldots
\left| 1\right> \left< 1\right|_n \nonumber \\
&+ & \left|\beta\right|^2 \left|\downarrow\right>
\left<\downarrow\right|
\left| 0\right> \left< 0\right|_1 \left|0\right> \left< 0\right|_2
\ldots
\left| 0\right> \left< 0\right|_n;.
\end{eqnarray}
Then the reliability of the records can be tested by looking for the 
basis in which all of the records are in accord. This can be accomplished
without destruction of all of the original redundant correlation between 
some of the records and the system.

These toy models establish that, in the presence of decoherence, it
is possible to record and that it is possible to find out what
information is reliable (which correlations are immune to
decoherence).  Again, we emphasize that the above toy models have
no pretense to even a remote kinship with real-world
observers.\footnote{Indeed, one could argue that if some unhappy
evolutionary mutation resulted in creatures which were bred to
waste their time constantly questioning the reliability of their
records, they would have become nourishment for other,
more self-assured creatures which did not need to pose and settle
such philosophical questions before making a useful prediction.}

\section{CONCLUSION}

What we have described above is a fairly complete sketch of the
physics involved in the transition from quantum to classical. 
Whether one would now claim that the emerging picture fits better
Bohr's ``Copenhagen'' framework or Everett's ``Many Worlds''
interpretation seems to be a semantic rather than a substantial issue. 
To begin with, decoherence was {\it not\/} a part of either of these
interpretations.  Thus, what we have presented here is clearly
beyond either CI or MWI.

The existential interpretation owes Bohr the central
question which was always implicit in the early discussions.  This
question --- about the location of the quantum-classical border --- is
really very similar to questions about ``existence.''  We have
posed and settled these questions operationally and, thus,
provided a quantum justification for some of the original CI program.

On the other hand, we owe Everett the observation that quantum
theory should be the key tool in the search for its
interpretation.  The question and concern may be traced to Bohr, but
the language of branches and the absence of explicit collapses and
{\it ab~initio\/} classicality are very much in tune with MWI.

We believe that the point of view based on decoherence settles
many of the questions which were left open by MWI and CI\null. This includes 
the origin of probabilities as well as the emergence of ``objective existence'',
although more needs to be done.

In particular, one issue which has been often taken for granted is looming big, 
as a foundation of the whole decoherence program.  It is the question of
what are the ``systems'' which play such a crucial role in all the discussions 
of the emergent classicality.  This issue was raised earlier,$^{2,28}$
but the progress to date has been slow at best.  Moreover, replacing
``systems'' with, say, ``coarse grainings'' does not seem to help at all 
--- we have at least tangible evidence of the objectivity of the existence of
systems, while coarse-grainings are completely ``in the eye of the
observer.''

It should be emphasized that reliance on systems does not
undermine the progress achieved to date in the study of the role of
decoherence and einselection.  As noted before,$^{11}$
the problem of measurement cannot be even stated without a recognition of the
existence of systems.  Therefore, our appeal to the same assumption
for its resolution is no sin.  However, a compelling explanation of what
are the systems --- how to define them given, say, the overall
Hamiltonian in some suitably large Hilbert space --- would be
undoubtedly most useful.

I would like to thank Chris Jarzynski, Michael Nielsen, and Chris Zalka for
comments on the manuscript.

\vfill
\eject

\begin{center}
\mbox{\psfig{figure=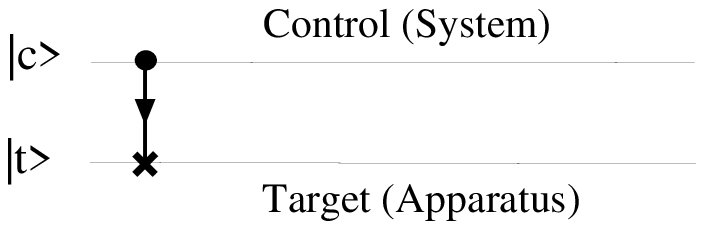,height=0.75in,width=2.0in}}
\bigskip
\begin{center}
\mbox{\psfig{figure=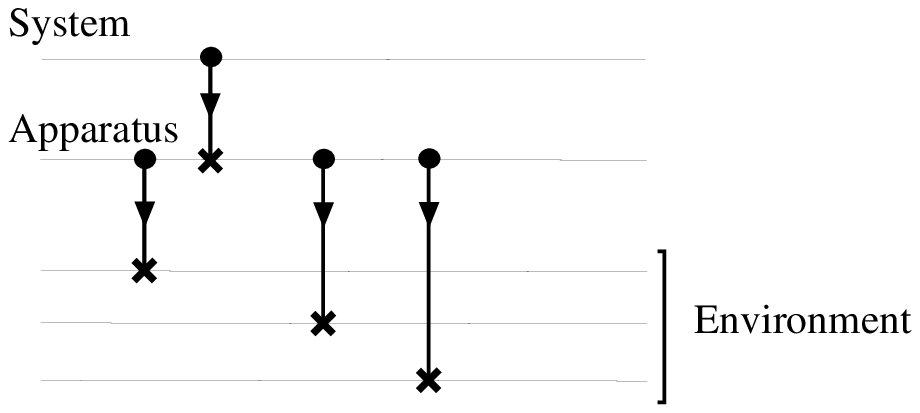,height=1.5in,width=3.2in}}
\end{center}\bigskip
\mbox{\psfig{figure=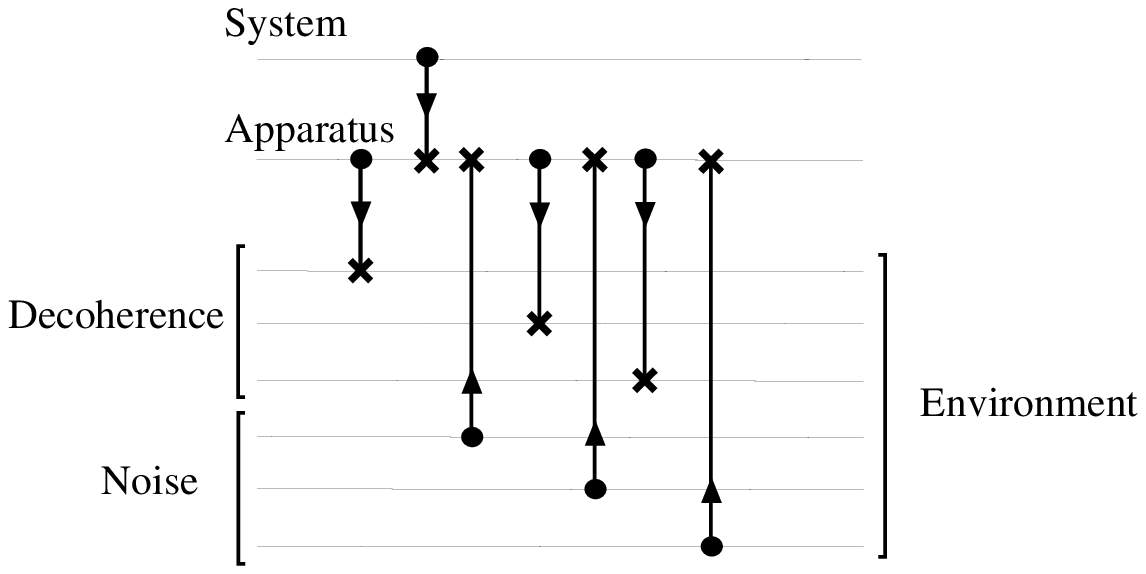,height=2.00in,width=4.0in}}
\end{center}
\noindent Fig. 1. Information transfer in measurements and in decoherence.

a) Controlled not ({\tt c-not}) as an elementary bit-by-bit measurement. Its
action is described by the ``truth table'' according to which the state of the
target bit (apparatus memory in the quantum measurement vocabulary) is 
``flipped'' when the control bit (measured system) is $|1>$ and untouched 
when it is $|0>$ (Eq. (2)). This can be accomplished by the unitary 
Schr\"odinger evolution (see, Refs. 1, 28, or 31 for the information theoretic
discussion).

b) Decoherence process ``caricatured'' by means of {\tt c-not}s. Pointer state 
of the apparatus (and, formerly, target bit in the pre-measurement, Fig. 1a)
now acts as a control in the continuous monitoring by the {\tt c-not}s of the
environment. This continuous monitoring process is symbolically ``discretized''
here into a sequence of {\tt c-not}s, with the state of the environment 
assuming the role of the multi-bit target. Monitored observable of the apparatus
-- its pointer observable -- is in the end no longer entangled with the system, 
but the classical correlation remains. Decoherence is associated with the 
transfer of information about the to-be-classical observables to the 
environment. Classically, such information transfer is of no consequence. In 
quantum physics it is, however, absolutely crucial, as it is responsible for 
the effective classicality of certain quantum observables, and for 
the relatively objective existence of preferred pointer states.

c) Noise is a process in which a pointer observable of the apparatus is 
perturbed by the environment. Noise differs from the purely quantum decoherence 
-- now the environment acts as a control, and the {\tt c-not}s which represent 
it carry information in the direction opposite to the decoherence {\tt c-not}s. 
Usually, both decoherence and noise are present. Preferred pointer observables 
and the associated pointer states are selected so that the noise is minimized.

\end{document}